\documentclass[11pt,twoside]{article}

\usepackage{asp2014}

\aspSuppressVolSlug
\resetcounters

\begin{document}

\title{Development of production-ready GPU data processing pipeline software for AstroAccelerate}

\author{Cees~Carels,$^1$ Karel~Ad\'{a}mek,$^1$ Jan~Novotn\'y,$^1$ and Wesley~Armour$^1$}
\affil{$^1$Oxford e-Research Centre, Department of Engineering Science, University of Oxford, 7 Keble Road, OX1 3QG, Oxford, United Kingdom}

\paperauthor{Cees~Carels}{}{0000-0002-8167-5717}{University of Oxford}{Oxford e-Research Centre (Department of Engineering Science)}{Oxford}{Oxfordshire}{OX1 3QG}{UK}
\paperauthor{Karel~Ad\'{a}mek}{karel.adamek@oerc.ox.ac.uk}{0000-0003-2797-0595}{University of Oxford}{Oxford e-Research Centre (Department of Engineering Science)}{Oxford}{Oxfordshire}{OX1 3QG}{UK}
\paperauthor{Jan~Novotn\'y}{}{0000-0002-9667-635X}{University of Oxford}{Oxford e-Research Centre (Department of Engineering Science)}{Oxford}{Oxfordshire}{OX1 3QG}{UK}
\paperauthor{Wesley~Armour}{wes.armour@oerc.ox.ac.uk}{0000-0003-1756-3064}{University of Oxford}{Oxford e-Research Centre (Department of Engineering Science)}{Oxford}{Oxfordshire}{OX1 3QG}{UK}



\begin{abstract}
Upcoming large scale telescope projects such as the Square Kilometre Array (SKA) will see high data rates and large data volumes; requiring tools that can analyse telescope event data quickly and accurately.
In modern radio telescopes, analysis software forms a core part of the data read out, and long-term software stability and maintainability are essential.

AstroAccelerate is a many core accelerated software package that uses NVIDIA\textsuperscript{\textregistered} GPUs to perform realtime analysis of radio telescope data, and it has been shown to be substantially faster than realtime at processing simulated SKA-like data.
AstroAccelerate contains optimised GPU implementations of signal processing tools used in radio astronomy including dedispersion, Fourier domain acceleration search, single pulse detection, and others.

This article describes the transformation of AstroAccelerate from a C-like prototype code to a production-ready software library with a C\texttt{++} API and a Python interface; while preserving compatibility with legacy software that is implemented in C.
The design of the software library interfaces, refactoring aspects, and coding techniques are discussed.
\end{abstract}



\section{Introduction}
Increasingly, there is a demand for faster than realtime data processing in radio astronomy.
In fact, it is a necessity for the planned SKA project; since the anticipated data volume and data rates would make storing all data for offline analysis unpractical.
For instance, low event-rate phenomena such as Fast Radio Bursts (FRBs) illustrate the need for efficient detection in realtime.

AstroAccelerate is a GPU-accelerated software library for analysing time-domain radio telescope data with NVIDIA\textsuperscript{\textregistered} Graphics Processing Units (GPUs) using CUDA\textsuperscript{\textregistered}.
AstroAccelerate's analysis modules consist of competitive implementations for dedispersion (\citet{armour2011gpubased}), single pulse search (SPS), Fourier domain acceleration search (FDAS) (\citet{Dimoudi_2018}), and others (\citet{admek2018gpu}, \citet{admek2019gpu}).

For example, SPS uses boxcar filters and peak-finding to maximise signal-to-noise, and detects pulses for SKA-like data faster than realtime (\citet{adamek2019single}).
To do this efficiently, telescope data are dedispersed and analysed entirely on the GPU.
In order to be production-ready, AstroAccelerate must be easy to configure, stable, and maintainable.
This paper describes the software considerations that led from a C-like prototype to a newly developed C\texttt{++} API that accomplishes this.

\section{Project Structure}
\subsection{Separate compilation and linking}
Before CUDA\textsuperscript{\textregistered} 5.0, linking could not be done separately from compilation.
In practice, it is desirable to compile host code with a separate host compiler such as gcc, even if this implies a modest decrease in performance due to fewer compile-time or linker optimisations.
For AstroAccelerate, CUDA\textsuperscript{\textregistered} code was refactored into separate source files with accompanying C-style wrapper functions declared in header files.

\subsection{Build systems and unit testing}
AstroAccelerate's build system was migrated from Make to CMake; simplifying configuration of CPU/GPU compiler flags according to the available architecture and compute capability.
Additionally, it is cumbersome to manually test memory allocation strategies on each GPU; so AstroAccelerate runs unit tests with CTest to check for proper allocation/de-allocation of GPU memory and graceful program exit.

\subsection{Memory management}
It is efficient to minimise GPU/host memory transfers between dedispersion and SPS; but it is cumbersome to parallelise (as well as debug) the code if each module relies on cudaMalloc() and cudaFree() having been used in a previous module.
Furthermore, each GPU has distinct capabilities.
Therefore, a memory management class was implemented.
Analysis modules can request GPU memory before they allocate it.
This reduces cluttering code with cudaGetLastError() and decreases the chance of memory leaks.
Other advantages of this approach are: 1. It decouples configuring a module from copying memory to the GPU, 2. It decouples dependencies of modules that run in sequence, 3. It facilitates parallel pipeline configuration/execution on multiple GPUs.

\subsection{Memory ownership}
Copying all memory from the user's control to the API is prohibitive.
The user always owns the telescope data memory on the host, and the API methods may not modify it.
Memory that the API requires on the GPU is encapsulated by the API.
Results data are transferred from the device to the host and this memory is owned and managed by the API; the API in turn provides access to the results data to the user.
The API uses const methods, const signatures, and returns by const reference or value as appropriate.

\section{API Structure}
\subsection{Design of the API}
The API offers a uniform pattern to configure and run analysis modules.
The user configures a C\texttt{++} \emph{plan} object for each module to set the desired search parameters and provide metadata.
The API constructs an accompanying \emph{strategy} object to determine an optimal configuration for the GPU (such as how input data are split and processed on the GPU).
A base class interface ensures each \emph{strategy} implements its own method to perform setup.
If there are configuration errors the API provides user feedback.
Internally, the API uses a std::map<std::set<\emph{component}>, bool> (\emph{component} is a strongly typed enum denoting analysis modules) to track which modules have been configured.
A schematic of the configuration pattern of the API is shown in Fig. \ref{fig:apischematic}.
The overall structure and underlying library code are shown in Fig. \ref{fig:apilogicalstructure}.

Using the API entails the following steps: 1. Provide telescope data, metadata, and select analysis modules (e.g. dedispersion, SPS, FDAS), 2. The API determines which \emph{pipeline} satisfies the selection of analysis modules, 3. Configure each module with a \emph{plan}, 4. Bind each \emph{plan} to the API, 5. For each \emph{plan}, the API calculates the optimal \emph{strategy}, 6. The API checks whether all modules are configured properly, 7. The API runs the \emph{pipeline}, 8. Transfer the results from the GPU to the host for the user.
\articlefigure[width=.80\textwidth]{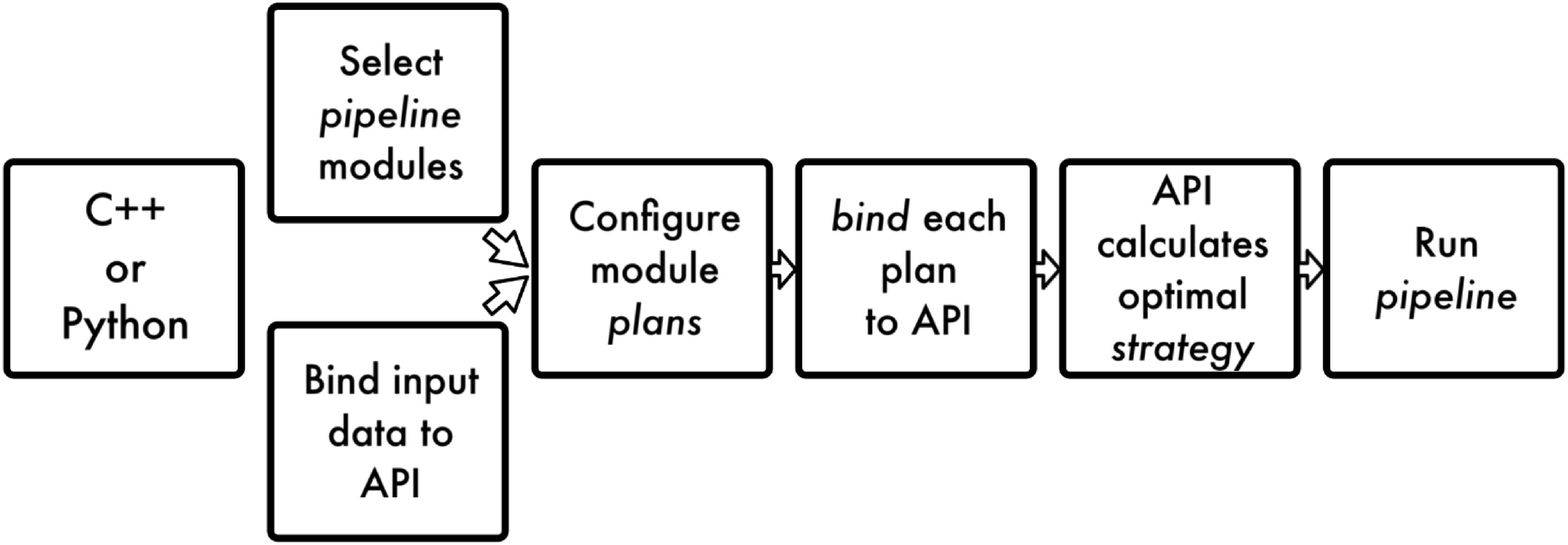}{fig:apischematic}{Schematic of the configuration pattern of the API.}
\articlefigure[width=.60\textwidth]{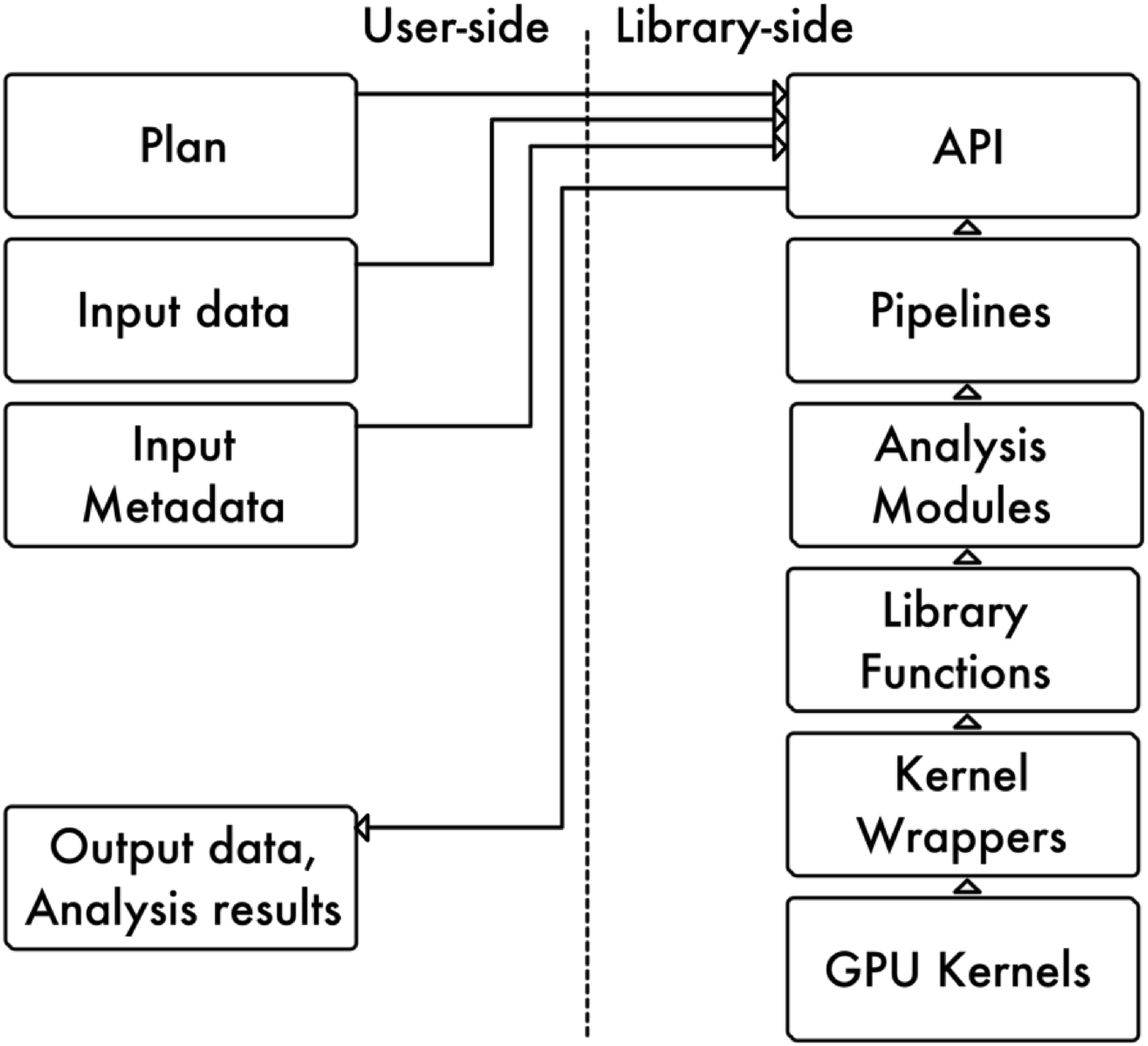}{fig:apilogicalstructure}{Overall structure of the API and the underlying library code.}
\subsection{\emph{Plan} and \emph{strategy}}
Several controls are imposed on the configuration of the API: 1. \emph{Strategy} objects are configurable through the API using a \emph{plan} and are immutable once configured, 2. The user may query the \emph{strategy} fields and ready-state, 3. \emph{Plan} and \emph{strategy} error checking are encapsulated, 4. The API informs the user which plans still need configuration, 5. The API ignores unnecessary \emph{plan}s.

\subsection{\emph{Pipeline}}
\emph{Pipeline}s are instances of \emph{pipeline} template specialisations.
Each template corresponds to a different combination of analysis modules.
Permitted pipelines have specialisations with constructor definitions (copy constructors are explicitly deleted): invalid configurations are compile-time errors.
All pipeline templates adhere to a base class interface; requiring methods to run the pipeline and obtain the results data.
The pipeline iterates over the radio telescope input data in chunks.
On each iteration, the user may query the pipeline via a \emph{status code}.
Pipeline status codes indicate whether the previous iteration was successful, if another module still needs to run, or if there is an error.

\section{Integration with Python}
A Python front-end delegates computationally intensive tasks to the C\texttt{++} API by calling into AstroAccelerate as a shared object library.
A Python user can direct the library to a file on disk, select analysis modules, configure \emph{plan} objects, execute a pipeline, obtain results, and query the pipeline status all from Python.
Using the Python front-end in this way and running dedispersion and SPS analogously to the standalone library did not show an appreciable runtime impact on the C\texttt{++} library \emph{pipeline} code.

\section{Conclusion}
The performance of the pipeline code was unaffected after implementing the API.
AstroAccelerate now has a modern C\texttt{++} API and a Python front-end which enable users to configure and run analysis pipelines to perform dedispersion, SPS, periodicity searches, and Fourier domain accelerated searches on radio telescope data using a GPU, and the AstroAccelerate library is now integrated into the SKA software repository.

The authors would like to acknowledge the use of the University of Oxford Advanced Research Computing (ARC) facility in carrying out this work.
\\https://doi.org/10.5281/zenodo.22558.
NVIDIA, the NVIDIA logo, and CUDA are trademarks and/or registered trademarks of NVIDIA Corporation in the U.S. and other countries.
Other company and product names may be trademarks of the respective companies with which they are associated.
\bibliography{P9-12}  
{\noindent \small \emph{ASP Conference Series in preparation.
Credit is hereby given to the ASP Conference Series.
\\\\This paper is covered by a publication agreement and copyright assignment with the Astronomical Society of the Pacific Conference Series (ASPCS).
\\\\No part of the material protected by this copyright notice may be reproduced or utilized in any form or by any means--graphic, electronic, or mechanical, including photocopying, taping, recording, or by any information storage and retrieval system--without written permission from the Astronomical Society of the Pacific.}}


\end{document}